# Exploiting ambipolarity in graphene field-effect transistors for novel designs on high-frequency analog electronics

*Francisco Pasadas[*], Alberto Medina-Rull, Francisco G. Ruiz, Javier Noe Ramos-Silva, Anibal Pacheco-Sanchez, Mari Carmen Pardo, Alejandro Toral-Lopez, Andrés Godoy, Eloy Ramírez-García, David Jiménez, and Enrique G. Marin[*]*

F. Pasadas, A. Medina-Rull, F. G. Ruiz, M. C. Pardo, A. Toral-Lopez, A. Godoy and E. G. Marin.
Departamento de Electrónica y Tecnología de Computadores, Pervasive Electronics Advanced Research Laboratory, Universidad de Granada, 18071 Granada, Spain.
E-mail: fpasadas@ugr.es, egmarin@ugr.es.

J. N. Ramos-Silva and E. Ramírez-García.
Instituto Politécnico Nacional, UPALM, Edif. Z-4 3er Piso, Ciudad de México, México.

A. Pacheco-Sanchez and D. Jiménez.
Departament d'Enginyeria Electrònica, Escola d'Enginyeria, Universitat Autònoma de Barcelona, 08193 Bellaterra, Spain.

Exploiting ambipolar electrical conductivity based on graphene field-effect transistors has raised enormous interest for high-frequency (HF) analog electronics. Controlling the device polarity, by biasing the graphene transistor around the vertex of the V-shaped transfer curve, enables to redesign and highly simplify conventional analog circuits, and simultaneously to seek for multifunctionalities specially in the HF domain. We present, here, new insights for the design of different HF applications such as power amplifiers, mixers, frequency multipliers, phase shifters, and modulators that specifically leverage the inherent ambipolarity of graphene-based transistors.

## 1. Introduction

Graphene is the most remarkable and striking member in the realm of two-dimensional materials. Just after its discovery, it was postulated as an optimal candidate for high-frequency (HF) electronic applications, thanks to its ultra-high, both, carrier mobility and saturation velocity[1], which are physically rooted to the massless-Dirac-fermions nature of the carriers on its distinctive band-structure. These exceptional transport properties have made graphene-based devices already competitive with state-of-the-art analog HF technologies[2], as evidenced by two main figures of merit related to the performance of HF transistors: the cut-off frequency, $f_T$, and the maximum frequency of oscillation, $f_{max}$. These frequencies set intrinsic upper limits to the operation of single de-embedded transistor-based amplifiers, according to two distinct drops in the gain: $f_T$ and $f_{max}$ are, respectively, the frequencies at which the current gain and power gain fall to unity.

Graphene-based field-effect transistors (GFETs) have already reached record $f_T = 427$ GHz,[3] close to maximum $f_T = 688$ GHz, exhibited by well-established III-V material-based High-Electron Mobility Transistors (HEMTs).[4] The performance of GFETs in terms of $f_{max}$ is not as impressive, but it is still remarkable as the highest $f_{max}$ value reported so far in GFETs is 200 GHz,[5] compared to several hundreds of GHz demonstrated by III-V competitors (even surpassing 1 THz in a InP HEMT device).[6] It is worth to note, however, that graphene is still a teenage material, while it is already offering competitive performance versus senior III-V semiconductors and HEMT architectures in the HF field.

Furthermore, graphene has proved excellent mechanical flexibility, anti-reflectance and corrosion resistance, and has thus become a main actor in the surge of nascent flexible and wearable nanoelectronics,[7–9] that will be of fundamental importance for the deployment of next generation HF ubiquitous wireless communication systems.[10] Importantly, it has been demonstrated to be compatible with thin-film technology, as well as with back-end-of-line processing, enabling its hybrid co-integration with commercial CMOS technologies in integrated circuits (IC).[11,12]

Enormous challenges, however, need still to be faced in graphene manufacturing technology, involving synthesis and substrate transfer (as they can severely impact on the carrier mobility via the interface quality), appropriate integration into monolithic millimeter-wave ICs, reproducibility and reliability. The better the achieved technological control, the closer will be graphene to higher technology readiness levels (TRLs) as well as to even more competitive HF figures of merit.

While much attention has been paid to the exploitation of outstanding graphene transport properties, its most distinct feature is, however, its inherent carrier ambipolarity, linked to its gapless electronic band-structure. This ambipolarity is exhibited through the V-shaped transfer characteristics (drain current *vs.* gate voltage) of GFETs around the point of minimum conductivity, namely the Dirac voltage,[13] enabling the electrical switching from hole to electron conduction and vice versa while operating.[14–16] This feature is opposed to unipolar silicon and III-V semiconductors-based devices, whose hole or electron conduction is predetermined during fabrication, through the chemical impurification of the semiconductor channel. Interestingly, since conventional electronics was completely founded on unipolar devices, this novel ambipolar feature of GFETs was initially assumed as undesirable and several ways were explored to suppress it.[17,18]

However, the ability to control the *operando* device polarity soon offered new design opportunities and, since the demonstration of the first realization of a GFET,[19] the ambipolarity of graphene has been leveraged in multiple applications. The most immediate was the frequency multiplication, achieved by electrically tunning the operation of the device around the Dirac point of the *I-V* parabola observed in GFETs transfer characteristics. This feature was exploited not only for signal generation at HF in several concomitant and early works,[20–25] but also for the development of high-performance biochemical sensing platforms[26] given the superior performance demonstrated in terms of 1/f and phase noise.[27] More advanced applications, directly raised by the graphene capability of frequency doubling, were rapidly conceived in terms of resistive subharmonic mixing,[28–33] which were lately demonstrated to be on par state-of-the-art performance.[2,34–36] A very particular application of GFET ambipolarity was demonstrated by X. Yang *et al*. in the form of a phase detector[37] by producing a DC output signal proportional to the phase difference of sinusoidal and square waves fed in the GFET gate terminal, while operating, again, at the vertex of the *I-V* parabola. Also exploiting the frequency doubling feature, the operation of frequency triplers[38] and quadruplers[39] have been demonstrated by cascading two GFETs to produce W-shaped transfer characteristics.[40] Furthermore, the non-linear operation caused by the ambipolarity has been leveraged in more sophisticated HF applications such as receivers,[41] and demodulators[42,43] achieving encouraging performance. The ambipolar conduction was also early interpreted as a tunable phase knot as demonstrated by the first realizations of modulators,[44] also conceived as (digital) ±180º-tunable power amplifiers.[45,46] Although the graphene ambipolarity was initially widely exploited in terms of parabolic *I-V* characteristics around the Dirac voltage, the gapless band structure was also exhibited in a V-shaped *C-V* curve of GFETs [47,48], and it was recently demonstrated to be suitable for phase-shifting purposes.[49] The ambipolarity distinct design opportunities have, indeed, extended beyond the analog arena into the digital domain, as demonstrated recently, e.g., in integrated security applications.[50]

In analog electronics, the keystone is, in any case, to exploit ambipolarity to conceive non-linear functionalities, enabling the precise tailoring of the main properties of HF signals: amplitude, frequency and phase. The controlled manipulation of these signal characteristics constitutes the basis of any HF wireless communication system. On this matter, this work brings new insights on leveraging the graphene ambipolarity to simplify the design of conventional applications and to seek out multifunctionalities from a device level, circuit design and an engineering perspective. We specifically analyze graphene-based systems that control (1) the amplitude in the form of power amplifiers; (2) the frequency in circuits such as mixers and frequency multipliers; and (3) the phase in applications such as phase shifters and phase-shift keying (PSK) modulators, paving the way towards the deployment of graphene-based ICs for HF wireless communications systems.

## 2. Circuit design of HF ambipolar graphene-based applications

The exploration of novel HF applications founded in ambipolar graphene transport requires a physics-based description of the electrical behavior of the GFET at an apt level to be exploited by circuit design suites. In particular, it becomes crucial to understand and model the electrical control of the GFET Dirac voltage ($V_{Dirac}$), as the threshold defining the different operation polarities of the transistor. For a well-behaved, i.e. ohmically contacted, long-channel GFET, and considering symmetric electron/hole mobility and carrier distributions, $V_{Dirac}$ is roughly given by:[51] $V_{Dirac} = V_{G0} + (V_D+V_S)/2$, where $V_{G0}$ is the offset voltage (embracing work-function differences between the gate metal and the graphene channel as well as the possible presence of additional charges due to impurities), and $V_D$ and $V_S$ are the intrinsic drain and source terminal biases, respectively, namely not affected by the potential drops at the metal contacts.[52] This rule of thumb, however, must be carefully considered in actual GFETs, where the electrical behavior is notably impacted by the metal-graphene contact resistances, with direct consequences also in the electrical control of $V_{Dirac}$. Although these contact resistances are in principle an undesired factor causing deleterious Dirac voltage shifts,[51] under proper engineering insights they can be harnessed as an additional degree of freedom for HF design. **Figure 1**a shows the equivalent resistive network of a GFET, including the contact resistances ($R_g$, $R_d$, and $R_s$, after gate, drain and source contacts) and the channel resistance ($R_{ch}$). $V_D$ and $V_S$ are determined by the external $V_{D,e}$ and $V_{S,e}$ and by $R_d$, $R_s$ and $R_{ch}$ as $V_D = V_{D,e}(R_{ch}+R_s)/R_t + V_{S,e}(R_d/R_t)$ and $V_S = V_{D,e}(R_s/R_t) + V_{S,e}(R_{ch}+R_d)/R_t$, where $R_t = R_{ch}+R_d+R_s$. The dependence of $V_{Dirac}$ on $R_d$ and $R_s$ is indeed a key enabler to the design of novel HF circuits and devices based on ambipolar transport. Of course, the value of these resistances should be as small as possible so to not spoil the GFET conductance, but thanks to the semi-metallic nature of graphene, and its electrically controllable work function, they are not fixed and can be tuned,[53] adding an extra degree of freedom to the already offered by the channel resistivity modulation (i.e. $R_{ch}$) in the control of $V_{Dirac}$.

However, the behavior of a GFET and the exploitation of its ambipolar nature cannot be reduced just to the understanding of the $V_{Dirac}$ dependencies and the modeling of the electrical behavior in terms of resistive elements. The current-voltage relations of a GFET operating at HF, and consequently the design of HF circuits based on its ambipolar response, requires of the description of the dynamic operation of the transistor. To this purpose, a large-signal model, able to deal with the intrinsic capacitive behavior of the GFET and applicable to e.g. transient, large-signal S-parameter (LSSP) or harmonic balance simulations, is needed. Although the exploitation of the ambipolarity of GFETs in HF circuit design must hold on this large-signal model (and as such has been considered in the different designs exploiting ambipolarity proposed later), very relevant insights can still be obtained from a small-signal equivalent circuit predicting the device behavior under AC or S-parameter simulations. This can be done, as far as the small-signal regime is applicable, in terms of a linear network of lumped elements. In **Figure 1**b, the charge-based small-signal equivalent circuit of the GFET (derived and validated against HF measurements of different GFET technologies)[54][55] is shown. The elements that conform the equivalent circuit are calculated from the voltage derivatives of both the current ($I_{DS}$) and terminal charges ($Q_G$, $Q_D$, and $Q_S$), by accounting for the charge conservation and non-reciprocity of the intrinsic capacitances of the GFET. Specifically, $g_m = dI_{DS}/dV_G$ is the transconductance, $g_{ds} = dI_{DS}/dV_D$ stands for the output conductance, and $C_{gs} = dQ_G/dV_S$, $C_{gd} = dQ_G/dV_D$, $C_{sd} = dQ_S/dV_D$, and $C_{dg} = dQ_D/dV_G$ are the four independent intrinsic capacitances of a three-terminal GFET. More details about the physics of these elements can be found elsewhere.[56]

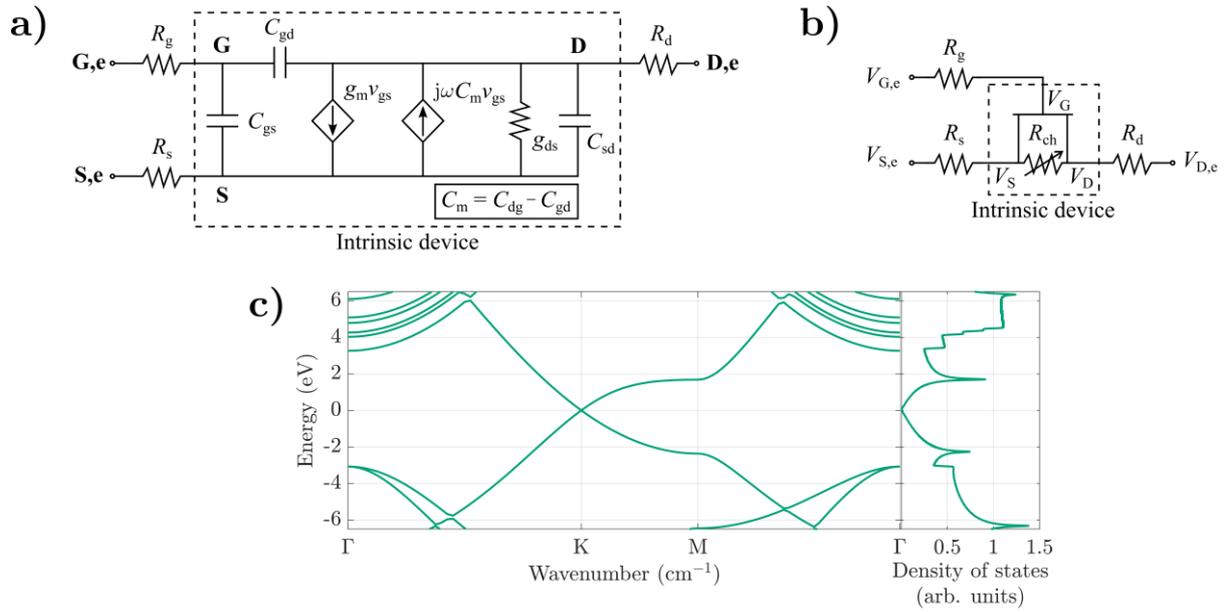

**Figure 1.- a)** Schematics of the GFET resistive network. The electrostatic modulation of the bias-dependent channel resistance ($R_{ch}$) is achieved via the gate terminal ($V_{G,e}$). Due to the non-negligible metal-graphene contact resistances ($R_s$ and $R_d$), a substantial potential drop is produced at both contacts. Thus, the intrinsic voltages ($V_S$ and $V_D$), not accessible in practice, are quite different from the external ones ($V_{S,e}$ and $V_{D,e}$). Reproduced from Ref. [51], CC licensing by 4.0, 2021. **b)** Charge-based small-signal model of a GFET. The small-signal elements are the transconductance $g_m$; the output conductance, $g_{ds}$; and the intrinsic capacitances $C_{gs}$, $C_{gd}$, $C_{sd}$, and $C_{dg}$. The physical meaning of the elements is thoroughly explained in Refs. [56–58]. $R_g$ is the gate resistance, and $R_d$ and $R_s$ account for the contact resistances of the drain and source, respectively. Reproduced with permission Ref. [55] Copyright 2017, IEEE.

Under this reduced but cardinal theoretical umbrella of the GFET, it is possible to devise novel HF circuits able to exploit the ambipolarity aiming to manipulate the amplitude, frequency and phase of HF signals. In the rest of the section, some modules for wireless HF communication systems that takes advantage of GFET-ambipolarity are exemplified. More specifically, we present designs that manipulate: the frequency (subsection 2.1), such as frequency multipliers and subharmonic downconverters; the amplitude (2.2), in the form of power amplifiers; and the phase (2.3), by addressing phase shifters, or modulators.

## 2.1. Frequency manipulation through graphene ambipolarity

**Frequency multipliers.-** The $V_{Dirac}$ shifting originated by tuning $V_D$ and $V_S$, and the graphene ambipolarity can be both exploited to build a circuit with a W-shaped transfer characteristic (TC), enabling frequency multiplication with control of the harmonic amplitudes.[40] The simplest implementation of such circuit considers a lumped resistor ($R_X$) between two GFETs connected in cascade. The resistor produces a controlled splitting between the $V_{Dirac}$ of both GFETs, $V_{Dirac,1}$ - $V_{Dirac,2}$, which is proportional to the voltage drop at $R_X$ (c.f. **Figure 2**a). Indeed, the role of $R_X$ can be played by the sum of extrinsic source and drain contact resistances of GFET1 and GFET2, respectively, as well as intentional graphene access resistance introduced between both devices. It is worth to note that the frequency multiplier design only demands a W-shaped TC and therefore the actual implementation could exploit different mechanisms for controlling the $V_{Dirac}$ splitting, e.g. with a single split-gate GFET architecture[59] or by means of different $R_s$ and $R_d$ metal-engineering in ultra-scaled structures.[53] Indeed, W-shaped responses have been already observed in back-gate GFETs,[40] although in an uncontrolled way as illustrated by the fact that devices with the same structure and dimensions presented different behaviors.

**Figure 2**a depicts the simplest design of a controlled frequency multiplier exploiting the GFET ambipolarity: the two GFETs are governed by the same gate bias ($V_{GG}$) and are connected in series through $R_X$. The drain of the first GFET is coupled to the power supply ($V_{DD}$) while the source of the second FET is DC grounded. The description of the employed graphene technology can be consulted in the Methods section. The output of the frequency multiplier is the GFET current. **Figure 2**b-c shows the resulting W-shaped TC for different values of $R_X$ (with $V_{DD}$ = 1V) and for several values of $V_{DD}$ (with $R_X$ = 1kΩ), respectively. The minima of the W-shaped TC correspond to the Dirac voltages of each GFET. The splitting between these charge neutrality points increases with the value of $R_X$, which also results in an increasing voltage drop at the resistor, reducing the overall current. The DC bias operation point of the multiplier completely defines the multiplication factor given by the frequency multiplier

(**Figure 2**d). Roughly, for tripler operation, $V_{GG}$ must be set in the mid-point between the $V_{Dirac,1}$ (or equivalently $V_{Dirac,2}$) and the relative maximum of the W, while for a frequency quadrupler, the operating bias point must correspond with the relative maximum (c.f. **Figure 2**d). It is well known that a frequency doubler can be also implemented just with one GFET by DC biasing at the $V_{Dirac}$, and this also holds for the cascaded design. The possibility of generating different harmonic multiplication factors simply by changing the DC operation point becomes a sort of reconfiguration capability enabled by the ambipolarity of the GFET, that substantially simplifies the circuit design for frequency manipulation. **Figure 2**d exemplifies the operation of the circuit as a frequency tripler and quadrupler with a sketch of input and output AC signals around the corresponding operation points (for $V_{DD} = 2V$ and $R_X = 1k\Omega$ with the TC shown with a green solid line in **Figure 2**c). A comparison between the predicted output signals (solid) and pure sinusoids at corresponding frequencies (dashed), shows there is still an evident distortion if the device is not optimized. As a case in point, for an input frequency of $f_{in} = 1MHz$, the analysis of the output power spectrum, achieved from circuit level simulations in the non-optimized scenario, indicates that the resulting HF relative power for the frequency tripler and the quadrupler at $3f_{in}$ and $4f_{in}$, respectively, are around the 50%. The performance of GFET-based frequency multipliers can nevertheless be improved by optimizing the W-shaped TC together with additional matching and stability networks.

**Subharmonic mixers.-** Ambipolar conduction can also be harnessed in the design of single GFET-based subharmonic mixers to exploit input frequency doubling before mixing by setting the transistor operation point at $V_{GS} = V_{Dirac}$. **Figure 3**a sketches the working principle: the local oscillator (LO) output ($v_{IN}$ in **Figure 3**a) feeds the gate terminal of the GFET. Leveraging on the quadratic response achieved by the V-shaped TC, the LO output frequency can be half of the actual mixing frequency ($v_{OUT}$ in **Figure 3**a). Subharmonic resistive mixers based on GFETs reaching state-of-the-art performance have already been demonstrated,[29,30,60] reducing the circuit complexity by only using one instead of two transistors and without the need of introducing two out-of-phase LO signals connected with a balun,[61] reducing the area and simplifying the fabrication process for eventual ICs.

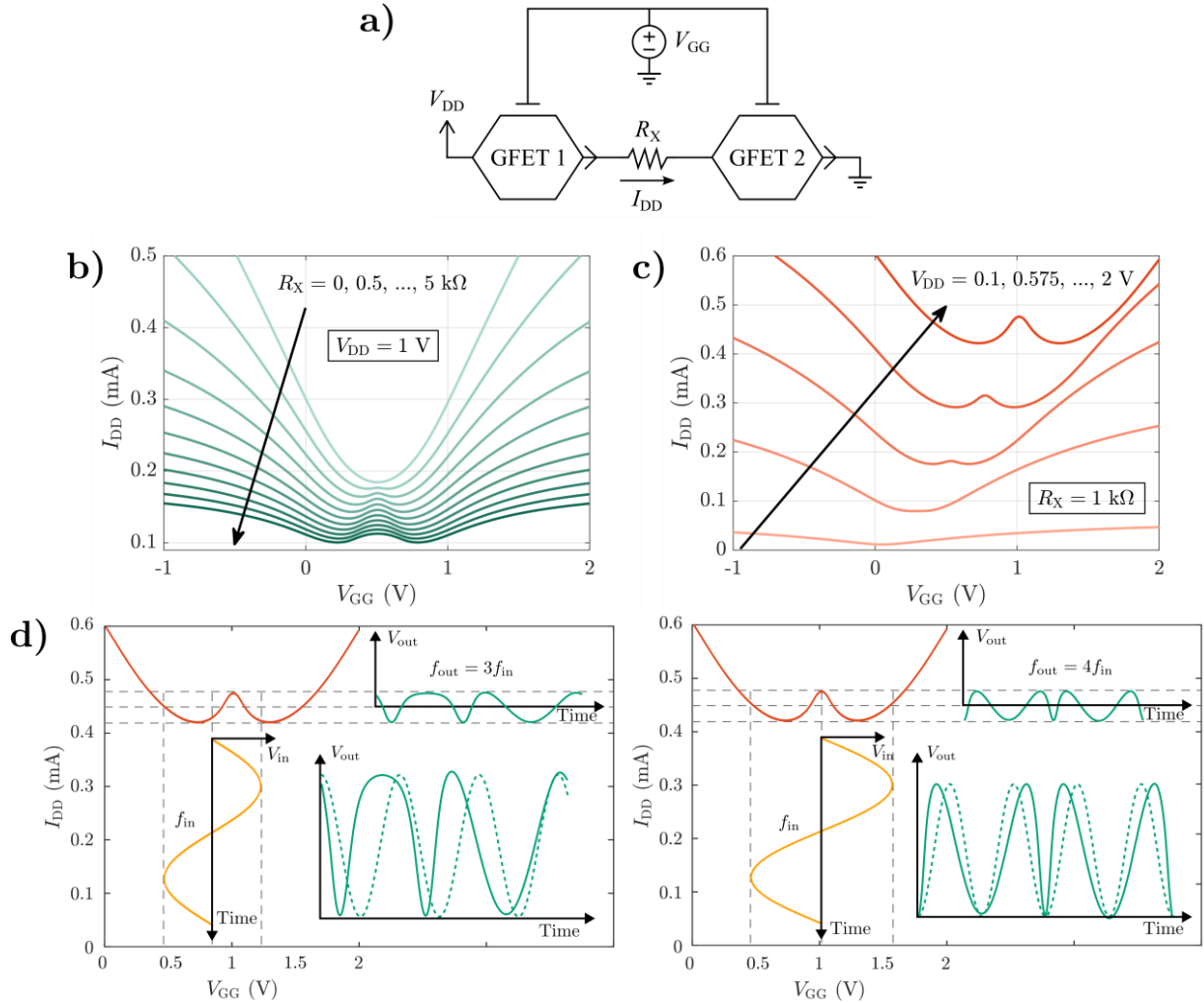

**Figure 2.- a)** Schematics proposed to build graphene-based frequency multipliers taking advantage of graphene ambipolarity. Each GFET is described in Table 1. The inclusion of a resistor, $R_X$, between two cascaded GFETs is proposed so to originate W-shaped transfer characteristics which can be leveraged to frequency multiplication.[51] Transfer characteristics for different **b)** values of $R_X$ (at $V_{DD}$ = 1V); and **c)** supply biases, $V_{DD}$ (considering $R_X$ = 1kΩ). **d)** Working principle for the GFET-based frequency tripler (left) and quadrupler (right) based on the TC at $V_{DD}$ = 2V and $R_X$ = 1kΩ. Inset: (green solid lines) Output signals of the frequency tripler (left) and quadrupler (right) compared against (green dashed lines) pure sinusoids at corresponding frequencies. Reproduced from Ref. [51], CC licensing by 4.0, 2021.

The subharmonic graphene-based mixer circuit is schematically shown in **Figure 3**b. The graphene technology considered is addressed in the Methods section. The radio-frequency (RF) signal ($f_{RF}$ = 2 GHz, $P_{RF}$ = -20 dBm) is connected to the drain, the LO signal ($f_{RF}$ = 1.01 GHz, $P_{LO}$ = -12 dBm) is connected to the gate through a bias tee, and the IF signal is collected at the IF port. The spectrum of the signal collected at the drain, with an IF output power of around -48 dBm, is shown in **Figure 3**c (blue solid lines) without filtering, as it was experimentally reported elsewhere (symbols).[29] If a perfectly quadratic *I-V* is assumed, negligible leakage of the LO signal can be expected at the drain terminal. However, to reach mixing operation at high frequencies, i.e. at the gigahertz band, significant gate-to-drain coupling can be expected through the GFET intrinsic gate-to-drain capacitance $C_{gd}$ (c.f. **Figure 1**b) and it becomes essential to filter out the LO frequency at the drain without affecting mixer output HF component. To do so, we included an open-loaded $\lambda_{LO}/4$ stub at the drain port in **Figure 3**b to effectively reduce the distortion of the mixed signal. This allows not only to short-circuit the fundamental LO frequency, but also (ideally) every odd multiple ($f_{LO}$, $3f_{LO}$, ...) because of the periodic behavior of the distributed element. On the contrary, at even-order components ($2f_{LO}$, $4f_{LO}$, ...), an open-circuit is generated. Thus, due to the proximity of the RF frequency and twice the LO frequencies ($f_{RF} \simeq 2f_{LO}$), RF remains almost unaffected. This aspect is critical to achieve that the maximum RF input signal reaches the transistor and participates in the mixing process.

The impact of the LO rejection filter is patently demonstrated in **Figure 3**c (red dashed lines), where the power spectrum at the drain of the GFET with the $\lambda_{LO}/4$ resonator is also depicted. A large attenuation is achieved for the main LO harmonic, and its odd-order multiples are greatly mitigated as well. These results show the great potential of not only leveraging graphene ambipolarity for subharmonic mixing, but also combining the different harmonics generated by the GFET quadratic TC by matching with distributed elements that also present a harmonically periodic behavior.

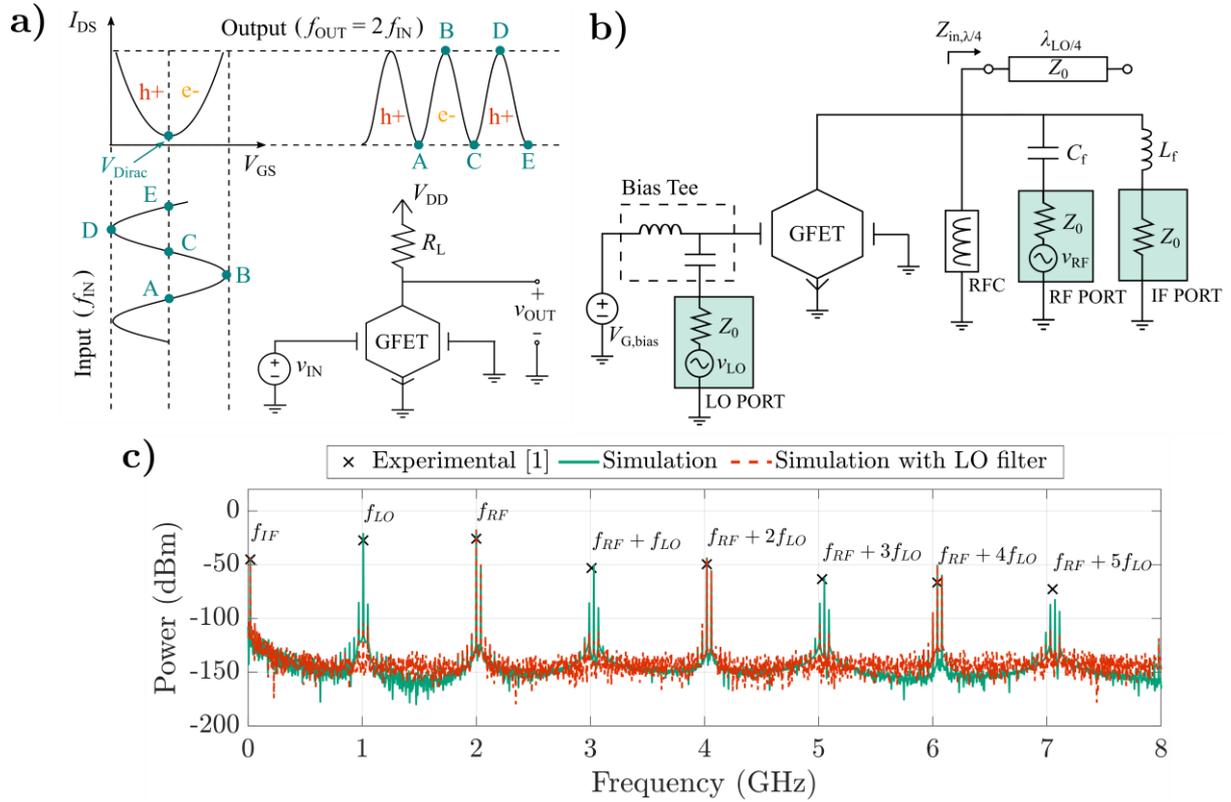

**Figure 3.- a)** Working principle of a GFET-based frequency doubler, relying on the inherent graphene ambipolar conduction. The same concept is leveraged in subharmonic mixers for generating a higher-frequency component at twice the LO frequency. **b)** Schematic showing the designed single-GFET subharmonic mixer. It includes the LO rejection filter consisting in an open-loaded $\lambda_{LO}/4$ stub. The RFC guarantees the operation at zero drain bias, while $V_{G,bias} = V_{Dirac}$. A bias tee is added to insert the LO signal, as well as $L_f$ = 265 nH and $C_f$ = 4 pF work as low- and high-pass filters, respectively. **c)** Frequency power spectrum of the signal collected at the drain of the GFET. Experimental data from Ref. [29] and simulation before (green solid line) and after (red dashed line) the addition of the LO rejection filter are compared.

*2.2 Leveraging ambipolarity for signal amplification*

**Power amplifiers.-** GFET ambipolarity can also be leveraged to efficiently amplify high-frequency signals and, simultaneously, providing symmetrical power gain under both electron and hole conductions, a unique feature to be employed in modulators[44] or high-frequency differential electronics applications such as baluns[45,46]. As amplification implies a linear processing of the signal, it is possible to apply microwave linear network analysis to the design of GFET-based power amplifiers.[62–64] To do so, we can make use of the equivalent circuit of the GFET (**Figure 1**b) in a two-port common-source configuration, to easily get the admittance matrix, $[Y_{int}]$, of the intrinsic device (framed in a blue box) and the contact resistance matrix, $[R]$, as:

$$[Y_{int}] = \begin{bmatrix} j\omega(C_{gd} + C_{gs}) & -j\omega C_{gd} \\ g_m - j\omega C_{dg} & g_{ds} + j\omega(C_{gd} + C_{sd}) \end{bmatrix} \quad [R] = \begin{bmatrix} R_g + R_s & R_s \\ R_s & R_d + R_s \end{bmatrix} \quad (1)$$

from which the global admittance matrix is obtained as: $[Y]^{-1} = [Y_{int}]^{-1} + [R]$. Using the relationships between two-port network parameters, the scattering matrix $[S]$ is obtained from $[Y]$ as:

$$[S] = \begin{bmatrix} \dfrac{(Y_0 - Y_{11})(Y_0 + Y_{22}) + Y_{12}Y_{21}}{\Delta Y} & -\dfrac{Y_{12}Y_0}{\Delta Y} \\ \dfrac{Y_{21}Y_0}{\Delta Y} & \dfrac{(Y_0 + Y_{11})(Y_0 - Y_{22}) + Y_{12}Y_{21}}{\Delta Y} \end{bmatrix} \quad (2)$$

where $Y_{ij}$ are the matrix elements of $[Y]$, and $\Delta Y = (Y_0 + Y_{11})(Y_0 + Y_{22}) - Y_{12}Y_{21}$, and $Y_0$ is the characteristic admittance.

It must be highlighted that $[S]$ describes a de-embedded device, i.e., after removing the extrinsic undesired contributions at HF produced by the interconnections and pads inherent to experimental prototypes. However, the metal-graphene contact resistances ($R_s$ and $R_d$), as would be present even in IC technology, are included in $[R]$. From $[S]$ it is possible to determine the maximum gain and microwave stability of GFET-based amplifiers. The maximum gain, $G_{max}$, is either the maximum available gain, $G_{MA} = |Y_{21}/Y_{12}|(K - \sqrt{K^2 - 1})$ (where $K = (2\Re(Y_{11})\Re(Y_{22}) - \Re(Y_{21}Y_{12}))/|Y_{21}Y_{12}|$) when the device is unconditionally stable; or the maximum stable gain, $G_{MS} = |Y_{21}/Y_{12}|$, when the device is potentially unstable. The rule of thumb in the design of amplifiers limits the operating frequency up to a 20% of $f_{max}$ to guarantee sufficient practical amplification.[65] In this frequency range, the vast majority of active devices become (without additional stability networks) potentially unstable, being $G_{MS}$ the gain that rules the design.

Thus, for the common-source configuration, $G_{MS}$ reads as

$$G_{MS} = \frac{1}{\omega}\sqrt{\frac{\omega^2(C_{dg} + R_s A)^2 + (g_m + R_s \omega^2 B)^2}{(C_{dg} + R_s A)^2 + R_s^2 \omega^2 B^2}} \qquad (3)$$

where $A = (C_{gd} + C_{gs})g_{ds} + C_{gd}g_m$ and $B = C_{gd}(C_{gd} + C_{gs} + C_{sd} - C_{dg}) + C_{gs}C_{sd}$. Interestingly, $G_{MS}$ is directly impacted only by the contact resistance at the common terminal, i.e., $R_s$ in the common-source configuration, although both $R_d$ and $R_s$ influence the actual value of the small-signal elements. If the contact resistance at the common terminal is negligible, i.e. $R_s \to 0$, the expression for the gain is simplified as:

$$G_{MS}|_{R_s \to 0} = \frac{\sqrt{C_{dg}^2 \omega^2 + g_m^2}}{\omega C_{gd}} \qquad (4)$$

namely $G_{MS}$ shows no dependence on $g_{ds}$. In other words, contrary to what would be expected from conventional unipolar technology arguments, the lack of current saturation is not a dominant factor straightforwardly impacting $G_{MS}$ in GFET amplification if the metal-graphene contact resistance is negligible, as it can be expected in more mature graphene technologies.

**Figure 4**a shows, $g_{ds}$, $|g_m|$ and $G_{max}$, as a function of $V_{GS,e}$ - $V_{Dirac}$, at a fixed $V_{DS,e} = 0.2$V (see Methods for material and device details). Shadowed in grey are the biases where $G_{max}$ falls below 0dB, and therefore the amplification is ineffective, while in green and yellow are the $V_{GS,e}$ biases where the GFET is unconditionally stable (US) and potentially unstable (PU), respectively.

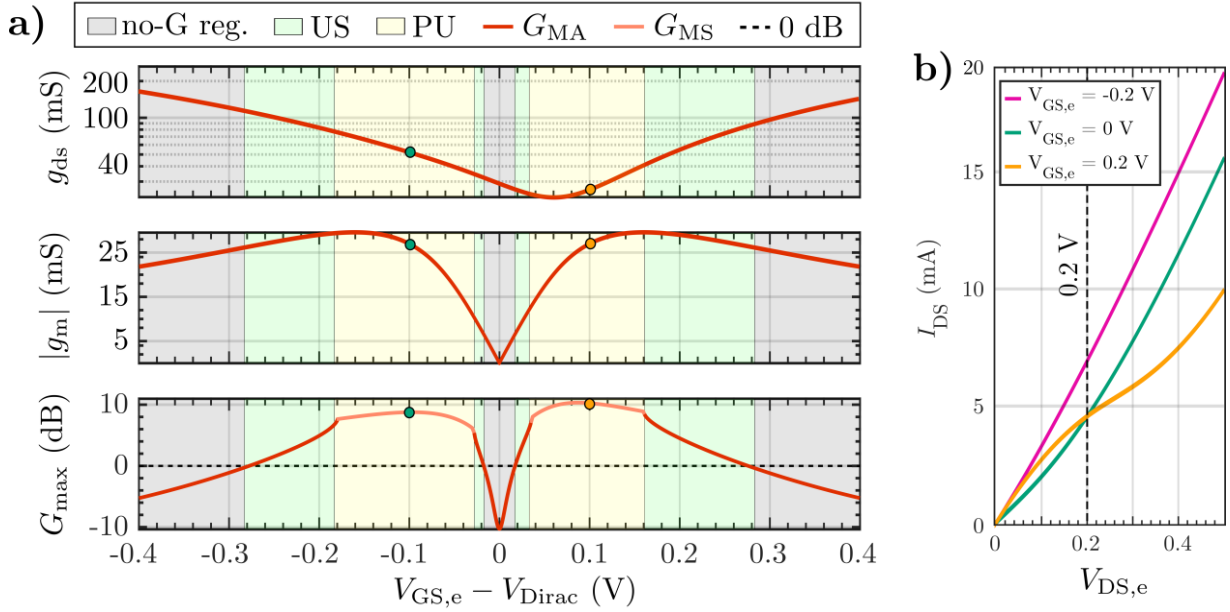

**Figure 4.- a)** Output conductance, $g_{ds}$; transconductance, $|g_m|$; and maximum gain, $G_{max}$, as a function of $V_{GS,e}$-$V_{Dirac}$, at a $V_{DS,e} = 0.2$V ($V_{Dirac} = 0.1$V). Legend: no-G: no gain region; US: unconditionally stable; PU: potentially unstable. **b)** Output characteristics of the GFET used for power amplification under three different $V_{DS,e}$ values: -0.2V, 0V, and 0.2V.

The $G_{max}$ maxima (red and black points) are found, for symmetric $V_{GS,e} - V_{Dirac}$ values, in both conductive branches of the ambipolar GFET. Notably, the highest $G_{max}$ value (red point; $V_{GS,e} > V_{Dirac}$) corresponds to a low $g_{ds}$ (actually close to its minimum value) while the second peak (black point; $V_{GS,e} < V_{Dirac}$) is correlated to a non-negligible value of $g_{ds}$, i.e. corresponding to a non-saturated region of the output characteristic ($I_{DS}$ vs. $V_{DS}$ curve). This can be clearly observed in **Figure 4**b, where the output curves are shown for both cases: $V_{GS,e} > V_{Dirac}$ (red point in **Figure 4**a and blue curve in **Figure 4**b); and $V_{GS,e} < V_{Dirac}$ (black point in **Figure 4**a and black curve in **Figure 4**b; the red curve in **Figure 4**b shows a case where $V_{GS,e} \ll V_{Dirac}$ showing a highly linear output characteristic). Both gain maxima are, however, located near the maxima of $|g_m|$, bringing to light that it is the transconductance, rooted in the graphene ambipolarity, the physical parameter that commands the power gain at HF –according to eq. (4)–. It should be also highlighted the symmetric behavior of $g_m$ around $V_{GS,e} = V_{Dirac}$ (in correspondence to symmetric hole-electron mobility) and the almost symmetric resemblance of $G_{max}$, in clear contrast to $g_{ds}$, which evidences the lack of a definite correlation between $G_{max}$ and $g_{ds}$. These conclusions prove that current saturation is not mandatory to achieve for power amplification based on GFETs thanks to theie inherent ambipolarity.

*2.3 Phase manipulation through graphene ambipolarity*

**Phase shifters.-** The last property of a signal that can be manipulated to process information is the phase. A phase shifter is the element specifically designed to tune the phase of an input signal according to a control signal, while the amplitude of the output is maintained either attenuated or amplified by a constant factor.[66] In a phase-controlled antenna array scenario, the phase shifter feeds each element of the array in a way that the amplitude and phase difference of the input current provided at each element determine the shape and direction of the main lobe of radiation, respectively. Thus, for a proper array design, it is of utmost relevance to be able to select the direction of the main lobe (changing the relative phases between the input signals of the antennas), while keeping the shape of the radiation pattern unaltered (maintaining the signal amplitudes of all elements balanced). GFETs are postulated as prime candidates for active phase shifters thanks to their bias-tunable quantum capacitance ($C_q$), related to the reduced density of states of graphene around the Dirac point, which is, once again, rooted into its particular semi-metal and ambipolar nature.[47]

To take advantage of this feature, the $C_q$ must be dominant over the gate insulator capacitance ($C_{ox}$) of the GFET. In particular, in the metal-insulator-graphene stack of a GFET, $C_{ox}$ and $C_q$ are associated in series,[48] and thus, $C_{ox}>>C_q$ is required to leverage the $C_q$ tunability in the overall capacitive response. A proper choice of $C_{ox}$ is therefore needed to guarantee a large modulation of the relevant intrinsic capacitances (see **Figure 1**b). **Figure 5**a-b, shows the variation of $C_{gs}$, $C_{gd}$, $C_{ds}$ and $C_{dg}$ with $V_{GS}$ and $V_{DS}$, for the technology parameters defined in the Methods. Notably, due to the graphene ambipolarity and strongly tunable occupation of the density of states, none of the intrinsic capacitances can be neglected in any bias region. This is, indeed, opposite to what happens in conventional semiconductor-based technologies where: i) in the subthreshold region, all intrinsic capacitances are irrelevant when compared with $C_{ox}$; and ii) in the saturation region, $C_{gd}$ can be obliterated given that the drain edge of the channel is depleted of mobile charged carriers.[67] Thus, in GFETs, any bias control signal through $V_{GS}$ and/or $V_{DS}$ can be eventually exploited for phase shifting operation through the variation of the capacitive response of the device.

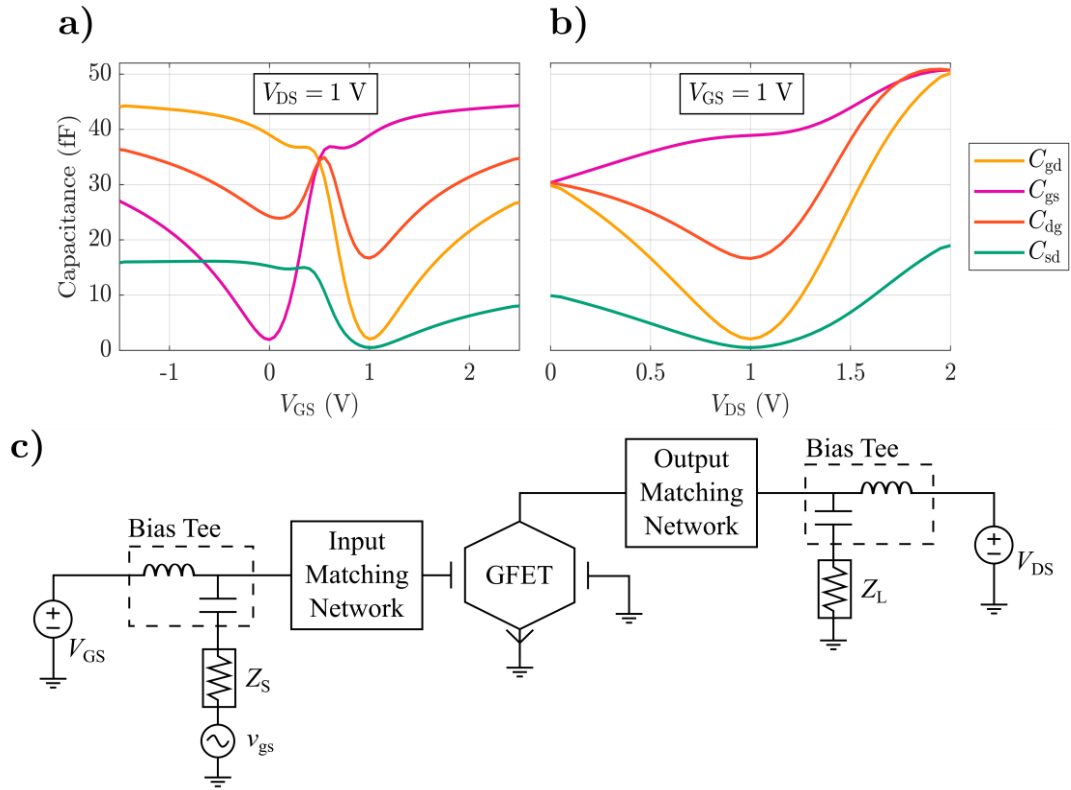

**Figure 5.-** Tunability of intrinsic capacitances $C_{gs}$, $C_{gd}$, $C_{sd}$, and $C_{dg}$ of the GFET employing the technology described in the Methods section versus **a)** gate bias, and **b)** drain bias. Reproduced from Ref. [66], CC licensing by 4.0, 2020. **c)** Schematics of a GFET-based phase shifter. The GFET is used as the active element. IMN and OMN allow to maximize the power transfer from the source to the load and, at the same time, minimize signal reflection from the load. Bias tees at both input and output ports are included, each one consisting of an ideal capacitor (inductor) to allow the AC (DC) through but uncoupling the DC (AC) signal. Reproduced from Ref. [66], CC licensing by 4.0, 2020.

To further exemplify the concept, **Figure 5**c shows the GFET operating in common-source configuration. The HF signal and DC control biases are combined using bias tees. Input and output matching networks (IMN and OMN, respectively) are employed to yield conjugate matching in both ports (see Methods).

In terms of the scattering parameters (cf. eq. (2)), a two-port phase shifter must be able to keep the magnitude of $S_{21}$ ($|S_{21}|$) constant while tuning its phase in a controlled way ($\phi_{21}$), where ports 1 and 2 of the system refer here to the gate-source and drain-source terminals, respectively. The rest of the $S$ parameters ($S_{11}$, $S_{12}$, and $S_{22}$) are also important to guarantee an acceptable power transfer from the input to the output and are addressed by the proper design of the IMN and OMN over a large window of $V_{GS}$ or $V_{DS}$ combinations (see Methods).

To get further insights of the $|S_{21}|$ and $\phi_{21}$ dependencies on the applied biases, **Figure 6**a-b depict their corresponding isocurves as a function of $V_{GS}$ and $V_{DS}$. As can be observed, both $|S_{21}|$ and $\Phi_{21}$ show a strong dependence on $V_{GS}$ and $V_{DS}$. It should be highlighted that each isocurve of **Figure 6**a-b provides a $V_{GS}$ – $V_{DS}$ combination ensuring a constant $|S_{21}|$ or $\phi_{21}$, respectively. Moreover, the phase isocurves depict a different dependence on $V_{GS}$ – $V_{DS}$ compared to amplitude isocurves, evidencing the possibility of applying a bias combination (i.e., a simultaneous variation of both $V_{GS}$ and $V_{DS}$) to yield a constant amplitude jointly with an appropriate phase variation. **Figure 6**c shows such $\phi_{21}$ variation (color scale) as a function of the bias combinations required to keep three constant values of $|S_{21}|$: -5 dB (dashed line), 0 dB (dotted line) and 5 dB (solid line). To ensure the unconditional stability of the circuit, the so-called $K - \Delta$ test is carried out and the results are included in **Figure 6**c by shadowing the regions in light grey and dark grey, corresponding to unconditional stability and conditional stability or instability, respectively. Hence, to provide an unconditional stable design, the circuit is limited to operate under the $V_{GS}$ – $V_{DS}$ combinations enclosed in the light grey region.

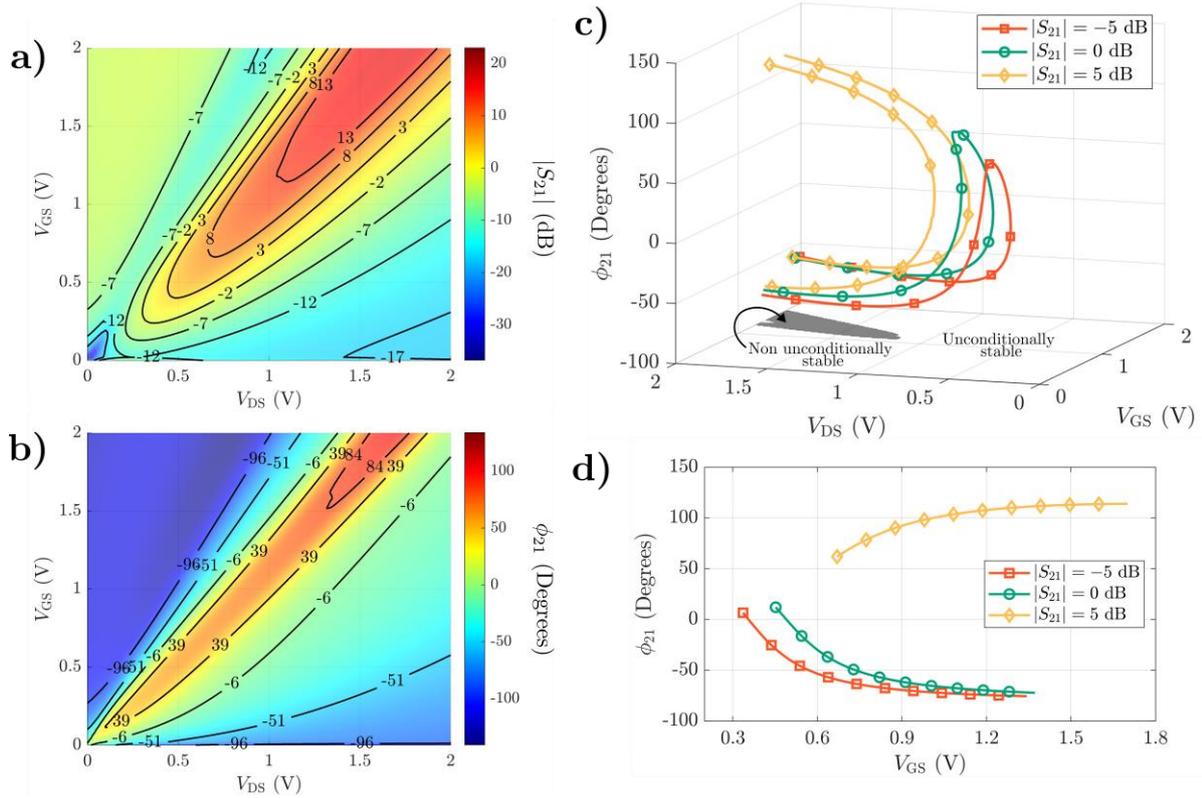

**Figure 6.-** Isocurve plots of **a)** $|S_{21}|$ (dB) and **b)** $\phi_{21}$ (Degree) versus both $V_{GS}$ and $V_{DS}$ for the GFET described in the Methods section at an operating frequency of 3GHz. **c)** Bias dependence of the phase shift $\phi_{21}$ for three gain values $|S_{21}|$ = -5 dB (orange line), 0 dB (green line) and 5 dB (yellow line). Bias combinations that do not guarantee unconditional stability for the device are colored in grey. **d)** Gate bias dependence of the phase shift for the same three constant gain values by considering that $V_{DS}$ follows a linear relation with $V_{GS}$ and thus it is simultaneously modified to maintain the chosen $|S_{21}|$ (analog control). Reproduced from Ref. [66], CC licensing by 4.0, 2020.

If any possible $V_{GS}$ – $V_{DS}$ combinations are allowed to keep a constant specific gain (i.e. digital control), that is to move arbitrarily along the amplitude isocurve, phase shifts as large as $\Delta\phi_{21} \cong 180°$ are achievable keeping $|S_{21}|$ = 0 dB. If a linear relation is forced between $V_{GS}$ and $V_{DS}$ (e.g. in a simple analog control implemented by a DC-DC converter or a voltage divider), the range of $\Delta\phi_{21}$ diminishes, with $\Delta\phi_{21}$ reaching values higher than 50° for the three gain values considered and with a remarkable $\Delta\phi_{21}$ higher than 80° for $|S_{21}|$ = 0 dB. These promising results turn GFETs into excellent candidates for future active phase shifter in HF communications systems.

**PSK modulators.-** GFET multifunctionality, due to ambipolar transport, enables the design of circuits, with one-single transistor, operating at different modes, and hence, enabling the development of area-efficient and robust applications, e.g., high-data rate modulators. In this realm, the ambipolarity- and bias-dependent functionalities of graphene transistors are specifically harnessed to design a 2.4 GHz phase-shift keying (PSK) modulator. The DC characteristics and the proposed multifunctional GFET circuit design are shown in **Figure 7**a-b, respectively.

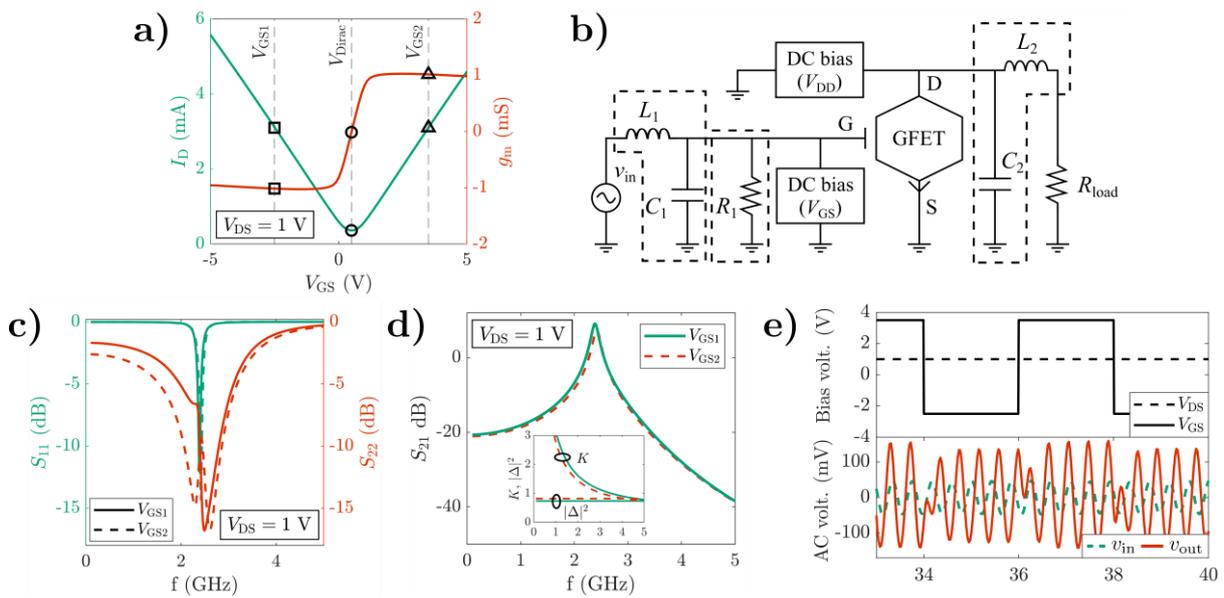

**Figure 7.- a)** Transfer characteristic (left axis) and corresponding transconductance (right axis) of the GFET with technology described in the Method section. The Dirac voltage is $V_{Dirac}$=0.5V (circle symbol) for $V_{DS}$=1V. The two selected gate voltages are highlighted $V_{GS1}$=$V_{Dirac}$-3V (square symbols) and $V_{GS2}$=$V_{Dirac}$+3V (tringle symbols). **b)** Schematic design of the proposed multifunctional GFET circuit. Elements inside the dashed limits are the matching (stability) networks. Filtering (DC/RF) blocks are not shown. $L_1$=98 nH, $C_1$=37 fF, $R_1$=72 kΩ, $L_2$=7.11 nH, $C_2$=476 fF. **c)** S-parameters ($S_{11}$ and $S_{22}$) and **d)** $S_{21}$ of the GFET-based multifunctional circuit at different $V_{GS}$. Inset in **d)** shows the stability parameters. **e)** DC bias voltage signals (top) and transient AC input and output voltage signals (bottom) of the GFET-based PSK modulator.

In-phase and inverting amplification functions have been obtained with the device biased within the *p*-type and *n*-type operation regimes, respectively,[16] e.g., at $V_{GS1}$=-2.5 V and $V_{GS2}$=3.5 V, i.e. gate biases of $V_{Dirac}$±3V. The high symmetry of the transfer characteristic of the GFET used here, with similar $I_D$ at $V_{GS1}$ and $V_{GS2}$ (cf. **Figure 7**a), facilitates the design of suitable IMN and OMN valid simultaneously for both bias points. Furthermore, unconditional stability conditions are also achieved for both $V_{GS}$ at a frequency of 2.4 GHz ($V_{DS}$=1V for both circuit operation modes). Importantly, the design is suitable for an on-chip hybrid implementation where the DC bias blocks can be implemented in a front-end CMOS technology while the GFET circuit corresponds to a back-end process. S-parameters and stability parameters of the multifunctional circuit are shown in **Figure 7**c-d for both operation modes at $V_{GS1}$ and $V_{GS2}$. The circuit (GFET + I/OMNs) is properly isolated at both the input and output ports since $S_{11}$ and $S_{22}$ are below -10 dB at 2.4 GHz. Furthermore, the multifunctional circuit also provides signal amplification since $S_{21}$ is 7 dB and 8.8 dB for the in-phase (at $V_{GS2}$) and inverting (at $V_{GS1}$) amplifiers, respectively.

The PSK modulation is demonstrated by the simultaneous application to the gate of the device of a baseband signal consisting of DC pulses varying from $V_{GS1}$ to $V_{GS2}$ with a data rate of 0.5 Gbps, and an input AC carrier $v_{in}$ of -15 dBm at 2.4 GHz while keeping $V_{DS}$=1 V. These input signals and the resulting output signal are shown in **Figure 7**d and, under such conditions, the circuit switches between operating modes, i.e., it works as an in-phase amplifier at $V_{GS2}$ and as an inverting amplifier at $V_{GS1}$, showing the unique ability of GFETs to perform as multifunctional circuits because of their inherent ambipolarity.

## 3. Conclusion

Ambipolar electronics based on graphene field-effect transistors have been demonstrated as ideal candidates to be exploited in the design of multiple high-frequency analog applications, showing the potential to simplify sophisticated circuit topologies even improving the circuit performance compared to more traditional approaches. This work gives insights about how the accurate control of graphene ambipolarity and other inherent features, such as the modulation of its quantum capacitance or the Dirac voltage tunability, can be exploited for the development of a variety of high-frequency electronic circuits such as power amplifiers, subharmonic mixers, analog phase shifters, frequency multipliers and high-data rate modulators. All these GFET-based circuits have proved advantages compared to well stablished implementations in terms of simplified designs, reduced number of components or reconfiguration capability. Thus, it can be expected that the relentless improvement of graphene manufacturing will overcome their present technological limitations providing the definitive momentum to the GFET-based analogue high-frequency electronic applications.

**Methods**

The design and analysis of HF applications founded in graphene field-effect transistors (GFETs) require a physics-based description of their electrical behavior at a compact and analytical level suitable for standard circuit simulators. To this purpose, we employ a large-signal model implemented in Verilog-A,[56,57] embedded it into Keysight Advanced Design System© (ADS). This GFET technology computer-aided design (TCAD) tool has been thoroughly validated[56,58] by the assessment of the DC characteristics, transient dynamics, and frequency response of a variety of graphene-based circuits such as a high-frequency voltage amplifier,[68] a high-performance frequency doubler,[21] a subharmonic mixer,[29] and a multiplier phase detector,[37] showing an excellent agreement between measurements and simulations. Although this work focuses on GFETs, some of the specific HF designs can be adopted for other ambipolar 2D technologies, such as phosphorene[69] or MoTe$_2$ FETs,[70] among others, e.g., by employing a suitable tool for circuit design based on these emergent ambipolar technologies such as the efficient algorithm for 2DFETs recently reported.[71]

In the following, we introduce the geometrical and technological TCAD parameters considered for each of the HF designs. The main parameters of the TCAD tool are: the effective carrier mobility, $\mu$; the gate offset voltage, $V_{G0}$; the residual carrier density, $n_{res}$; the channel width and length, $W$ and $L$, respectively; and the oxide capacitance per unit area, $C_{ox}=\varepsilon_{ox}\varepsilon_0/t_{ox}$, where $\varepsilon_{ox}$ is the relative permittivity of the gate oxide, $\varepsilon_0$ is the vacuum permittivity, and $t_{ox}$ is the gate oxide thickness.

*Graphene technology electrical and physical TCAD parameters for the frequency multiplier*
In the case of the frequency multiplier, we consider a standard graphene technology consisting in a 10nm $Al_2O_3$ ($\varepsilon_{ox}$=9) gate stack, with graphene dimensions 1µm×1µm. In our approach, the role of the quality of the graphene channel is played by the mobility and the residual charge density.[72,73] In this regard, a normal quality of graphene is considered by assuming a mobility of 2000cm$^2$/Vs and a residual carrier density of 6.86×10$^{11}$cm$^{-2}$ at room temperature. The offset voltage is 0V, a variation of this parameter in an eventual fabricated technology would mean a shift on the operating gate bias. Finally, a contact resistivity of 200Ω×µm and a gate resistance of 5Ω are assumed.

*Graphene technology electrical and physical TCAD parameters for the subharmonic mixer*
The first realization of a subharmonic mixer was reported in Ref. [29]. The technology consisted of a monolayer graphene sheet generated using the method of micromechanical exfoliation applied to natural graphite. The sample was positioned onto a 300-nm silicon oxide film that was grown through thermal process on a high-resistivity silicon substrate. Given the size constraints imposed by the exfoliated graphene flake, a gate length of 1 µm and width of 20 µm (2 fingers × 10 µm) was arranged. Electron beam lithography was employed to define the drain/source pads, consisting of a layer stack comprising Ti (1 nm), Pd (30 nm), and Au (70 nm) metals. For the top-gate dielectric, a natural oxidation process was employed to form a 2-nm-thick layer of Al, followed by the deposition of a 25-nm layer of $Al_2O_3$ ($\varepsilon_{ox}$=9) through e-gun evaporation. An electron beam lithography technique was then used to create a top-gate electrode, which was composed of Ti (1 nm), Pd (30 nm), and Au (60 nm). The reported mobility was 2000 cm$^2$/Vs and 2400cm$^2$/Vs and a residual carrier density of 10$^{12}$cm$^{-2}$ was extracted.[29] In this regard, for the technological prediction of the sub-harmonic mixer performance we assume $\mu=2200$ cm$^2$/Vs and $n_{res}=9.22\times10^{11}$cm$^{-2}$. The offset voltage was measured in 1V and the contact resistivity was experimentally estimated in 560Ω×μm while a gate resistance of 21Ω is considered.

Regarding the zero-bias mixer topology shown in **Figure 3**b, the RF signal ($f_{RF}$ = 2 GHz, $P_{RF}$ = -20 dBm) is introduced to the drain of the GFET through a high-pass filter (HPF), the LO signal ($f_{RF}$ = 1.01 GHz, $P_{LO}$ = -12 dBm) is connected to the gate through a bias tee, and the IF signal is collected at the IF port with a low-pass filter (LPF). Both first-order filters are designed with the cut-off frequencies of 800 MHz and 30 MHz, respectively, resulting in ideal lumped components of $C_f$ = 4 pF and $L_f$ = 265 nH, respectively.

*Graphene technology electrical and physical TCAD parameters for the power amplifier*
The graphene technology considered for the power amplifier design fits (not shown in this work) the reported technology developed by the Institut d'Electronique de Microélectronique et de Nanotechnologie (CNRS, Lille, France) reported elsewhere and described in the following.[54,74–76] The device fabrication starts with the application of e-beam lithography on a 300 nm SiO$_2$/highly resistive Si substrate to pattern the gate. Next, a 40-nm layer of Al was deposited and lifted off. To finish the gate structure, a natural oxidation process was employed, where the substrate was exposed to air for 24 hours, leading to the formation of a 4.3-nm layer of Al$_2$O$_3$, as confirmed by spectroscopic ellipsometry. Such natural oxidation process eliminates the need for high-temperature deposition of a dielectric layer, making it compatible with the fabrication of GFETs on flexible substrates. The graphene was grown using chemical vapor deposition (CVD) on a copper foil, resulting in a 1.5 cm × 1.5 cm area. A wet chemical transfer process based on polymethyl methacrylate (PMMA) was employed to transfer the monolayer graphene sheet onto the pre-patterned gate. To obtain isolated patterns, reactive ion etching (RIE) with oxygen plasma was used to etch the graphene channel to (width × length) 48µm×0.388µm. Source and drain contacts were defined by depositing a 20 nm layer of Ni followed by a 30 nm layer of Au and subsequently performing a lift-off process. To enable compatibility with on-chip probe measurements, the device fabrication concluded with the deposition of a 50 nm/300 nm Ni/Au stack for the gate, source, and drain pads. A resulting carrier mobility of 3000cm$^2$/Vs and a residual carrier density of 3.55×10$^{11}$cm$^{-2}$ at room temperature were obtained. The offset voltage is 1V, the contact resistivity is 480Ω×µm, while a gate resistance of 25Ω is assumed

*Graphene technology electrical and physical TCAD parameters for the analog phase shifter and the PSK modulator*

In the case of the PSK modulator and phase shifter designs, we consider a standard graphene technology but using a gate oxide that satisfies $C_{ox}$=110fF/µm$^2$ >> $C_q$ in order to leverage the GFET quantum capacitance ($C_q$) tunability, especially for the latter application. Similar to the technology considered for the frequency multiplier design, the assumed graphene dimensions are 1µm×1µm; a normal quality of graphene is considered (mobility of 2000cm$^2$/Vs and a residual carrier density of 6.86×10$^{11}$cm$^{-2}$ at room temperature); and the offset voltage is 0V. The contact resistivity is 100Ω×µm in the case of the phase shifter and a state-of-the-art value of 10Ω×µm is considered in the case of the PSK modulator.[77] The gate resistance is 5Ω (25Ω) for the phase shifter (modulator) circuit.

Regarding the phase-shifter circuit in amplifier topology shown in **Figure 5**c, the HF signal and DC biases are combined by using bias tees and, to achieve a good power transfer, two matching networks are employed. Eventually a shunt resistor of 1.65 kΩ (not included in the topology) is added to the gate of the GFET to increase its stability at the cost of some gain losses. Unconditional stability is achieved for $V_{GS}$ and $V_{DS}$ = 1V, which allows to calculate the reflection coefficients $\Gamma_s$ and $\Gamma_L$ for the maximum available gain $G_{MA}$. Input and output matching networks (IMN and OMN, respectively) are designed to yield conjugate matching in both ports resulting in both cases in a shunt capacitor ($C_{IMN}$ = 465 fF, $C_{OMN}$ = 55 fF) and a series inductor ($L_{IMN}$ = 35 nH, $L_{OMN}$ = 37 nH). The IMN is configured in a C-L topology while the OMN is configured in a L-C topology. It is important to note that the lumped components are assumed to be ideal. Tolerance and quality factor (Q) issues associated with a realistic implementation could affect the design and, therefore, their influence should be analyzed in detail according to the integrated circuit technology employed in an eventual realization of the circuit. Additionally, the matching networks are designed for an operating frequency of 3 GHz and, generally, for a single bias point, but in the case of the phase-shifter application, both $V_{GS}$ and $V_{DS}$ of the GFET have to be varied to enable the phase shifting while keeping a constant amplitude, hence, it is crucial to achieve a high matching coefficient $(M)$[78] value for a large window of $V_{GS}$ or $V_{DS}$ combinations. In this regard, we have assessed that $M$ is over 0.7 for the bias window $V_{GS}$, $V_{DS}$ = [0, 2] V.

**Acknowledgements**

This work is funded by FEDER/Junta de Andalucía-Consejería de Transformación Económica, Industria, Conocimiento y Universidades through the Projects A-TIC-646-UGR20, B-RNM-375-UGR18 and P20_00633; by Junta de Andalucía-Consejería de Universidad, Investigación e Innovación under ENERGHENE Project no. P21_00149; and by MCIN/AEI/10.13039/501100011033 through the projects PID2020-116518GB-I00 and PID2021-127840NB-I00 (MCIN/AEI/FEDER, UE). We also acknowledge the support by the European Union's Horizon 2020 Framework Programme for Research and Innovation through the Project GrapheneCore3 under Grant Agreement No. 881603. F. Pasadas acknowledges funding from PAIDI 2020 – European Social Fund Operational Programme 2014–2020 no. 20804. A. Medina acknowledges the support of the MCIN/AEI/PTA grant, with reference PTA2020-018250-I. A. Pacheco-Sanchez acknowledges the support from from Ministerio de Ciencia, Innovación y Universidades under grant agreement FJC2020-046213-I. M.C. Pardo acknowledges the FPU program (FPU21/04904). A. Toral-Lopez acknowledges Plan Propio programme from Universidad de Granada. E. Ramírez-García acknowledges the support from IPN Contract no. SIP/20230362.

[10.1002/smll.202303595](10.1002/smll.202303595)